\documentclass[prl,twocolumn,showpacs,preprintnumbers,superscriptaddress]{revtex4} 
\usepackage{graphicx} 

\usepackage{dcolumn} 
\usepackage{color} 
 
\newcommand{\beq}{\begin{equation}} 
\newcommand{\eeq}{\end{equation}} 
\newcommand{\beqa}{\begin{eqnarray}} 
\newcommand{\eeqa}{\end{eqnarray}}

\begin{document} 
 
\title{From infinite to two dimensions through the functional renormalization group} 
\author{C.~Taranto} 
\affiliation{Institute for Solid State Physics, Vienna University of Technology 1040 Vienna, Austria} 
\author{S.~Andergassen} 
\affiliation{Faculty of Physics, University of Vienna, Boltzmanngasse 5, 1090  
Vienna, Austria} 
\author{J.~Bauer}
\affiliation{Department of Physics, Harvard University, 17 Oxford St.,  MA 02138, USA}
\author{K.~Held} 
\affiliation{Institute for Solid State Physics, Vienna University of Technology 1040 
Vienna, Austria} 
\author{A.~Katanin} 
\affiliation{Institute of Metal Physics, Russian Academy of Sciences and Ural Federal University, Ekaterinburg, Russia}
\author{W.~Metzner} 
\affiliation{Max Planck Institute for Solid State Research, 70569 Stuttgart, Germany}
\author{G.~Rohringer}
 \affiliation{Institute for Solid State Physics, Vienna University of Technology 1040 Vienna, Austria} 
\author{A.~Toschi} 
\affiliation{Institute for Solid State Physics, Vienna University of Technology 1040 Vienna, Austria} 
 
\begin{abstract} 
We present a novel scheme for an unbiased, 
non-perturbative treatment of strongly correlated fermions.
The proposed approach combines two of the most successful many-body methods, the  dynamical mean field theory (DMFT) and the functional renormalization 
group (fRG). 
Physically, this allows for a systematic inclusion of non-local 
correlations via the fRG flow equations, after the local correlations are taken into account non-perturbatively by the DMFT. 
To demonstrate the feasibility of the approach, we present numerical results for the
two-dimensional Hubbard model at half-filling.  
\end{abstract} 
 
\date{\today} 
\maketitle

{\sl Introduction.} -- Correlated electron systems display undoubtedly some of the most
fascinating phenomena of  condensed matter physics such as
high-temperature superconductivity and quantum criticality; and 
with the tremendous progress to cool and control  
atomic gases new many-body physics is  explored nowadays. These systems
pose a particular challenge for theory. 
In this paper, we discuss a new route for the theoretical treatment
of strong correlations, which combines the 
strengths of two of the most successful approaches developed hitherto: 
 dynamical mean field theory (DMFT) \cite{DMFT,DMFT2}  and 
functional renormalization group (fRG) \cite{Berges,Renormalization,Kopietz,fRGreview}.

DMFT represents the ``quantum'' extension of 
the classical (static) mean-field theory \cite{DMFT2}. More formally, it provides the exact solution of a quantum many-body Hamiltonian  in
 the limit of infinite spatial dimensions ($d \rightarrow \infty$)\cite{DMFT}. 
DMFT allows hence for an  accurate (and non-perturbative) treatment of
the {\sl local} part of the correlations. Among others, it 
provides the essential ingredients to describe the 
Mott-Hubbard metal-to-insulator transition in three-dimensional bulk systems \cite{DMFTreview,V2O3review}. 
At the same time, the mean-field nature with respect to the spatial degrees of freedom implies
that all {\sl non-local} spatial correlations are completely
neglected in DMFT.

A powerful technique to treat such {\sl non-local} correlations is,
instead, provided by the fRG. Its starting point is an exact 
functional flow equation  \cite{Wetterich93}, which yields the gradual evolution
from a simple initial action
to the full final action, that is, the generating functional of all
one-particle irreducible vertex functions. The flow parameter (RG
scale) is usually a momentum or energy cutoff. Expanding the
functional flow equation yields an exact but infinite hierarchy of 
flow equations for the $n$-particle vertex functions, which for 
most calculations is truncated at the two-particle level.
There have been many applications of such weak-coupling 
truncations to low-dimensional fermion systems with competing 
instabilities and non-Fermi liquid behavior (for a review, see \cite{fRGreview}).

The approach we present here is coined DMF$^2$RG as 
the DMFT solution serves as a starting point of the fRG flow.
DMF$^2$RG aims at overcoming the main restrictions 
of the two methods, i.e., the lack of non-local correlations in DMFT
and the weak-coupling limitation in practical implementations of the fRG. The basic idea of the DMF$^2$RG 
is the following: We apply the fRG not starting from a problem without (or with trivial) correlations, but from a converged DMFT solution of the 
correlated system. This way, the local but possibly strong DMFT
correlations, essential to capture the Mott-Hubbard physics, are fully taken into account from the very beginning.
{\sl Non-local} correlations beyond DMFT, particularly important for low-dimensional systems, will be systematically generated by the fRG flow.  
We note that alternative strong coupling starting points for the fRG flow were recently discussed for the Bose-Hubbard \cite{dupuis} and the single-impurity Anderson model \cite{Kinza}.
 
Before turning to the DMF$^2$RG algorithm, let us mention alternative approaches proposed in the past to include non-local correlations beyond DMFT.
They can be classified into cluster \cite{cluster,cluster_rev}
and diagrammatic extensions \cite{DGA,DF,1PI,FLEX,Multiscale2,Multiscale} of DMFT. The
former ones are evidently complementary in nature to DMF$^2$RG,
as they provide {\sl short}-range correlation beyond DMFT, but at a high numerical cost, which poses
significant limits to multiband calculations.
Similarly as the diagrammatic extensions of DMFT, the DMF$^2$RG includes short- and long-range correlations on equal footing and improves the scaling with the
number of orbitals. At the same time, instead of a simple selection of diagrams (e.g. second order perturbation theory, ladder, etc.), DMF$^2$RG
 exploits the more powerful RG and generates 
parquet-like diagrammatic corrections to DMFT. 
This way, DMF$^2$RG provides a systematic and {\sl unbiased} treatment of electronic correlations beyond DMFT in {\sl all channels}.
Topologically the same diagrams albeit with different Green's functions and vertices are obtained in the proposed parquet implementations of D$\Gamma$A \cite{DGA} and multi-scale methods \cite{Multiscale2,Multiscale}.
This is however computationally much more demanding, and suffers from divergences of the two-particle irreducible vertex \cite{Multiscale,Schaefer2013,Janis2014}.

\begin{figure}[tb] 
\begin{center} 
\includegraphics[width=8cm]{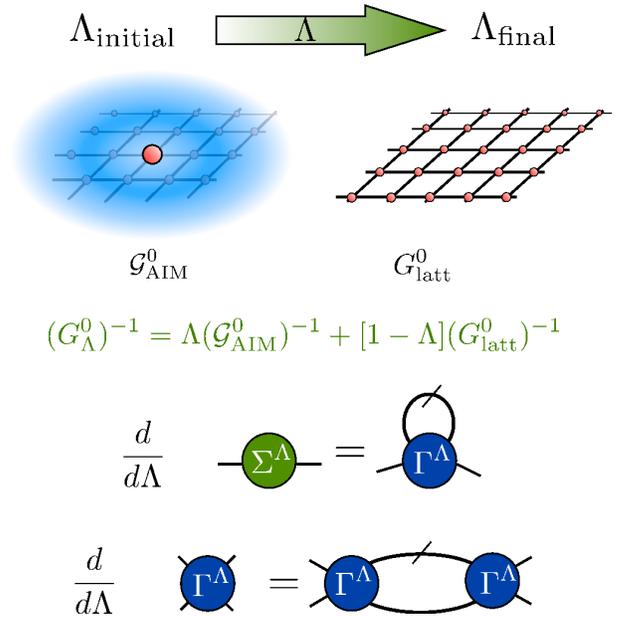} 
\end{center} 
\par 
\vspace{-.5cm} 
\caption{(Color online) Schematic illustration of the DMF$^2$RG approach, showing the evolution of the Gaussian part $G^0_\Lambda$ of the action from DMFT to its exact expression for a two-dimensional system.  The (truncated) flow equations for the self-energy $\Sigma^{\Lambda}$ and  the two-particle vertex $\Gamma^{\Lambda}$  are explicitly given in terms of Feynman diagrams.} 
\label{Fig1} 
\end{figure}

{\sl Method.} -- 
A rather flexible and effective formulation of DMF$^2$RG (see also the Supplementary Material section for further details) is obtained
starting from the local (or ``impurity'') action of DMFT
\begin{eqnarray}
\mathcal{S}_{\mathrm{DMFT}} &=&-\!\int_0^\beta\! \!d\tau d\tau' \sum_{i\sigma} \bar{c}_{i\sigma}(\tau) \mathcal{G}_{\mathrm{AIM}}^0(\tau-\tau')^{-1}c_{i\sigma}(\tau') \nonumber\\
&&+ \; \mathcal{S}_{int} \;.
\label{eq:S_DMFT}
\end{eqnarray}
Here, $\bar{c}_{i\sigma }$($c_{i\sigma }$) are the Grassmann variables corresponding to the creation
(annihilation) of a fermion with spin orientation $\sigma=\uparrow, \downarrow$  on site $i$, $\mathcal{G}_{\mathrm{AIM}}^0(\tau-\tau')$ is the electronic-bath Green's function  of the auxiliary  effective Anderson impurity model (AIM), which in a first step  
needs to be determined  self-consistently in DMFT \cite{DMFTreview} (see left-hand side of Fig.\ \ref{Fig1}), and $\mathcal{S}_{int}$ is a local interaction.

With this DMFT solution as a starting point, 
the fRG  generates a flow to the finite-dimensional action of interest
\begin{eqnarray}
\mathcal{S}_{\mathrm{latt}} &=&-\int_0^\beta\!\!d\tau d\tau'
\sum_{\mathbf{k}\sigma}
\bar{c}_{\mathbf{k}\sigma}(\tau)G^0_{\mathrm{latt}}(\mathbf{k},\tau-\tau')^{-1}c_{\mathbf{k}\sigma}(\tau')\nonumber\\
&&+ \;\mathcal{S}_{int},
\label{eq:S_latt}
\end{eqnarray}
where $G^0_{\mathrm{latt}}(\mathbf{k},\tau-\tau')$ is the free propagator of 
the finite dimensional system, which reads
$G^0_{\mathrm{latt}}(\mathbf{k},i \omega)= (i\omega -
\epsilon_{\bf k} +\mu)^{-1} $ in terms of Matsubara frequencies,
the energy-momentum dispersion $\epsilon_{\bf k}$ and the chemical
potential $\mu$.  In Fig.\ \ref{Fig1} the specific case of a $2D$ square lattice is shown. 

For the  DMF$^2$RG scheme we now introduce a flow parameter $\Lambda$ \cite{note1} so that
$
G_\Lambda^0(\mathbf{k},i\omega)^{-1}=\Lambda
  \mathcal{G}_{\mathrm{AIM}}^0 (i\omega)^{-1}+ (1-\Lambda) G^0_{\mathrm{latt}}(\mathbf{k},i\omega)^{-1},
\label{eq:cutoff}
$
interpolates between the initial DMFT ($\Lambda_{\rm initial}=1$) and the final action ($\Lambda_{\rm final}=0$).

The flow of DMF$^2$RG hence 
gradually switches off the DMFT-bath and switches on the 2D hopping, including non-local correlations beyond DMFT. 
Neglecting  three (and more) particle vertices, the flow equations \cite{fRGreview,Salmhofer2001}
for the self-energy and the two-particle vertex are shown in Fig.~\ref{Fig1}.
The truncation of the hierarchy at the level of the two-particle vertex $\Gamma$ relies on the assumption that the relevant physics is captured by the structure appearing on the two-particle level. Let us emphasize, however, that three- (and more-) particle vertices are included on the local level by DMFT.
This flow scheme results in the following single-scale propagator (defined as ${\partial G_\Lambda}/{\partial \Lambda}|_{\Sigma^\Lambda \mbox{fixed}}$)
\begin{equation}
S_\Lambda(\mathbf{k},i\omega)\!=\! G^2_\Lambda(\mathbf{k},i\omega)\!
\left[G^0_{\mathrm{latt}}(\mathbf{k},i\omega)^{-1 } \!-\!\mathcal{G}^0_{\mathrm{AIM}}(i\omega)^{-1} 
\right]
\label{eq:single}
\end{equation}
which includes the full Green's function $G_\Lambda(\mathbf{k},i\omega)=[G_\Lambda^0(\mathbf{k},i\omega)^{-1}-\Sigma^\Lambda(\mathbf{k},i\omega)]^{-1}$.

While the formal structure of the flow equations, diagrammatically
depicted in Fig.~1, resembles the one
 of the conventional fRG implementation, in the DMF$^2$RG
the initial conditions strongly differ, as they are determined, both
at the one- and the two-particle level, 
by DMFT, which provides the initial
self-energy $\Sigma^{\Lambda=1}= \Sigma_{\mathrm{DMFT}}(i\omega)$ and
one-particle irreducible (1PI) 
vertex  $\Gamma^{\Lambda=1} =
\Gamma_{\mathrm{DMFT}} (i\nu_1, i\nu_2;i\nu'_1,i\nu'_2)$ \cite{notevertex}. 
As a consequence, DMF$^2$RG is numerically  more expensive
than the conventional fRG or DMFT schemes:
(i)  two-particle vertices have to be computed in DMFT \cite{Rohringer2012} as an input to the 1PI-fRG flow 
and (ii) the frequency dependence of $\Sigma^\Lambda$ and
$\Gamma^\Lambda$ has to be included in the fRG \cite{Kinzanew}, with a
proper frequency-dependent parametrization; according to a generic estimate the numerical effort scales as $N_k^{4} N_\omega^{4}$, $N_k$($N_\omega$) being the number of momenta (frequencies). DMF$^2$RG allows to bypass the sign-problem of a direct quantum Monte Carlo (QMC) treatment of non-local correlations, since QMC will be limited, at most, to DMFT calculations of one- and two-particle local vertices.

\begin{figure}[tb] 
\begin{center} 
\includegraphics[width=8cm, clip=true]{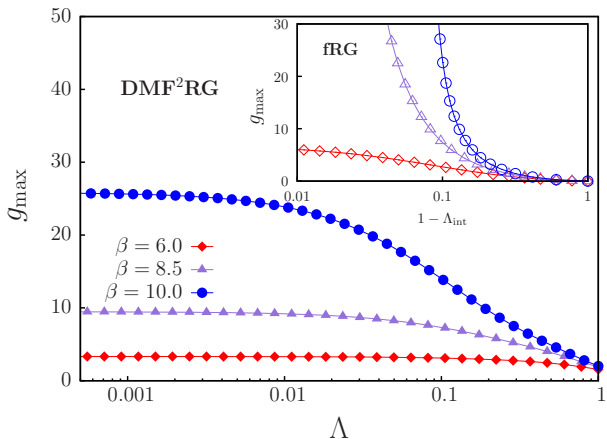} 
\end{center} 
\vspace{-.5cm} 
\caption{(Color online) Flow of the largest component ($g_{\rm max}$) of the two-particle vertex function, i.e., in our case, $\Gamma$ in the particle-hole crossed channel, for zero transfer frequency ($\nu_2-\nu'_1=0$), antiferromagnetic momentum transfer (${\bf k}_2-{\bf k}'_1= (\pi,\pi)$) and ${\bf k}_1=(0,\pi)$, ${\bf k}_2 = (\pi,0)$ computed by fRG, with interaction cutoff $\Lambda_{\rm int}$\cite{HRAE2004} (inset) and DMF$^2$RG (main panel) for the two-dimensional half-filled Hubbard model at $U=1$, at different (inverse) temperatures.} \label{Fig2} 
\end{figure}

{\sl Application to the 2D Hubbard Model.} -- 
 We now show, as a first application of 
DMF$^2$RG, results for a prototypical model of correlated fermions,
the two-dimensional Hubbard model. 
We recall that the interplay of antiferromagnetism and superconductivity
in this model has been studied by weak coupling
truncations of various versions of the fRG already some
time ago \cite{Ref1,Ref2,Ref3,Ref4}.
In standard second-quantization notation, the Hubbard Hamiltonian reads \cite{Hubbard} 
\begin{equation} 
H=-t\sum_{\langle ij\rangle \sigma }c_{i\sigma }^{\dagger }c_{j\sigma 
}+U\sum_{i}n_{i\uparrow }n_{i\downarrow }  \label{H} ,
\end{equation}%
where $t$ denotes the nearest-neighbor hopping amplitude on a square
lattice and $U$ the local
Coulomb repulsion. In the following,  we will define our
energies in terms of $4t\equiv 1$, and fix the average  particle density
to half filling
$n=1$.  In this case, 
the momentum transfer of $(\pi, \pm \pi)$
corresponds to perfect (antiferromagnetic) nesting on the square shaped Fermi surface.

We solve the truncated flow equations numerically, including the self-energy feedback in the equation for $\Gamma^\Lambda$. We use a channel decomposition of the interaction vertex \cite{Karrasch2008,supple} with Matsubara frequency dependence of the self-energy and the interaction vertex. 
The  momentum-dependence is taken into account by discretizing the Brillouin zone into patches with constant coupling function. If fine enough, this discretization captures the angular variation of the coupling function along the Fermi surface with good precision.
For simplicity, we restrict ourselves to $8$ patches, which already includes important physical aspects of the $2D$ system \cite{gull2010}.

\begin{figure}[t!] 
\begin{center} 
\includegraphics[width=8cm, clip=true]{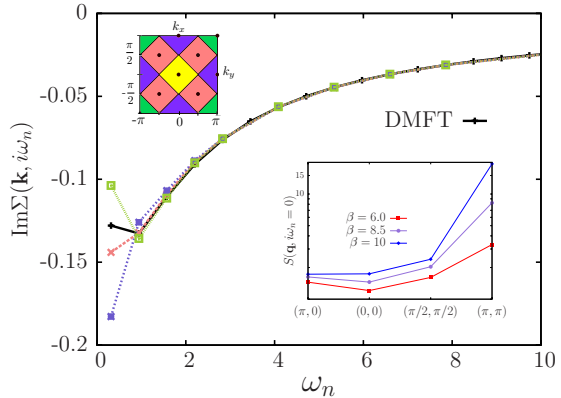} 
\end{center} 
\par 
\vspace{-.5cm} 
\caption{(Color online) Comparison of the results for the imaginary
  part of the fermionic self-energy of the two-dimensional Hubbard
  model for $U=1$, and $\beta= 10$, calculated within DMFT ({\bf
    k}-independent, in black) and DMF$^2$RG, for different ${\bf
    k}$-vectors (the color coding of the different ${\bf k}$ is
  defined in the inset, note that the values of Im$\Sigma({\bf k},
  i\omega_n)$ for ${\bf k}=  (0,0) $ and $(\pi,\pi)$ coincide because
  of the particle-hole symmetry). {\sl  Upper inset}: Scheme of the
  $8$-patches discretization used for the calculations. {\sl Lower
    inset}: $T-$dependence of the momentum-resolved spin correlation function $S({\bf q},i\Omega=0)$. }
\label{Fig3} 
\end{figure} 

\begin{figure}[tb] 
\begin{center} 
\includegraphics[width=8cm]{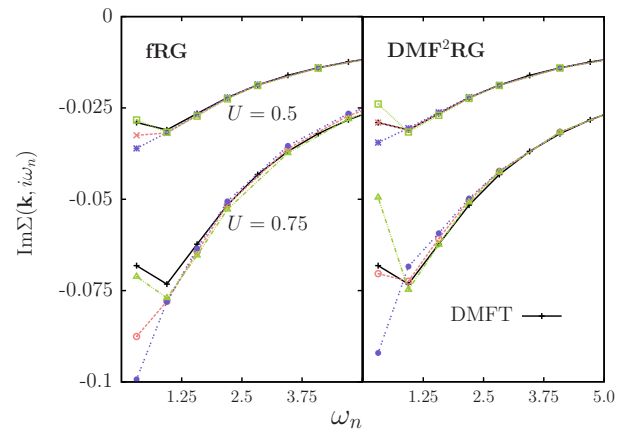} 
\end{center} 
\par 
\vspace{-.5cm} 
\caption{(Color online) Comparison of the imaginary part of the  self-energy for $U= 0.5, 0.75$, $n=1$, and $\beta=10$, calculated by fRG and DMF$^2$RG, for different ${\bf k}$-vectors (color coding as in Fig.~\ref{Fig3}).} 
\label{Fig4} 
\end{figure}

{\sl Numerical results.} -- Our calculations for the
two-particle vertex function and self-energy are reported in
Figs.~\ref{Fig2} and \ref{Fig3}-\ref{Fig4},
respectively. In Fig.~\ref{Fig2} we plot the largest
component ($g_{\rm max}$) of the vertex function, which -- at
half-filling -- is found in the particle-hole crossed channel for zero
frequency and antiferromagnetic momentum transfer ($\pi,\pi$). The
data, which refer to a weak-intermediate regime ($U \!=\! 1$),
clearly show that the DMF$^2$RG mitigates the fRG tendency to a
low-$T$ divergence of the flow: We still obtain a converged  DMF$^2$RG result for $g_{\rm max}$ at $\beta=1/T=10$ 
, whereas the fRG flow for the vertex is manifestly divergent \cite{note_cutoff}. Quantitatively, by fixing
an upper-bound for $g_{\rm max}$, we observe that the temperature at which it is reached is slightly decreased in DMF$^2$RG compared to fRG for moderate values of the interaction (up to $U=0.75$) while is significantly decreased from $T\! \sim \!0.125$ 
(fRG) to $\!\sim \!0.085$ (DMF$^2$RG) at $U=1$. This is attributed to the
damping effect of the local correlations, included from the very beginning in the flow of DMF$^2$RG. We emphasize that this ``divergence'' is {\sl not} associated with a true onset of a long-range order. In fact, fRG-schemes can be adapted to access also the disordered phase at lower $T$ \cite{Wetterich2004}, though such an extension goes beyond the scope of this work.

We now turn to the analysis of the self-energy results obtained with
the DMF$^2$RG flow at the lowest temperature considered, i.e.,
$\beta=10$. Here, the fRG flow diverges, and it is worth to compare
the DMF$^2$RG results with the original DMFT data, see
Fig.~\ref{Fig3}. As expected in $2D$, the non-local
correlations captured by the DMF$^2$RG  strongly modify the DMFT
(${\bf k}$-independent) results, determining a significant momentum
dependence of the self-energy at low frequencies: While in DMFT a
metallic solution, with a moderate Fermi-liquid renormalization of the
quasi-particle mass,  is obtained, in DMF$^2$RG we observe a strong
enhancement of the imaginary part of the self-energy at the Fermi
surface.  In fact, at the ``antinodal'' point ($\pi$, $0$), where the
largest value of $-$Im$\Sigma$ is found, the low-frequency behavior is
manifestly non quasi-particle-like, indicating the destruction of the
Fermi surface in this region of the Brillouin zone.
The trend of  large non-local corrections to DMFT at the antinodal momentum and towards a pseudogap formation is similar to cluster-DMFT results \cite{cluster_rev,Olle_priv}.  
Deviations from the DMFT metallic results, albeit less marked, are
found at the ``nodal'' point ($\frac{\pi}{2}$, $\frac{\pi}{2}$), for
which one cannot exclude, at this temperature, a residual presence of
strongly damped quasi-particle excitations. The significant reduction
of  $-$Im$\Sigma$ w.r.t. DMFT, observed at
($0$,$0$) or ($\pi$,$\pi$), does not imply  metallicity since these
points are far away from the Fermi surface; and the real part of the
self-energy (not-shown) is also strongly enhanced w.r.t.\ DMFT.
A further insight on the non-local correlations captured by the
DMF$^2$RG is given by the analysis of the momentum/frequency-dependent
susceptibilities, which in DMF$^2$RG can be extracted from the two-particle vertex.
In the lower inset of Fig.\ \ref{Fig3}, we show the DMF$^2$RG results for the
momentum-resolved spin-susceptibility at zero frequency $S({\bf q},  i\Omega\!=\!0)$. This quantity is most important at half-filling, where  magnetic fluctuations predominate, and it is experimentally accessible, e.g., via neutron spectroscopy. Our results are in qualitative agreement with the QMC data of Refs.\ \cite{White1989,Bickers1991} and show the major role played by antiferromagnetic fluctuations, with a pronounced peak at $(\pi,\pi)$, growing upon decreasing $T$. 
 The ferromagnetic fluctuations also get enhanced due to the van Hove singularity at the Fermi level.

In Fig.~4, we  compare the DMF$^2$RG self-energy data with the fRG.
 The comparison can only be performed at weaker coupling
 and/or higher $T$ than in Fig.~\ref{Fig3}, as the fRG flow needs to converge.
Our numerical data of Fig.~\ref{Fig4} indicate that in the considered
parameter region (same $T$, but weaker interaction than in
Fig.\ \ref{Fig3}) the fRG and DMF$^2$RG yield qualitatively similar
results for the {\bf k} dependent self-energy. Considering that in
DMF$^2$RG {\sl local} correlations have been included
non-perturbatively via DMFT, this confirms the validity of previous
fRG analysis of the Hubbard model at weak and moderate interaction.  At the same time, the applicability of DMF$^2$RG goes beyond the weak-to-intermediate coupling of the fRG, allowing for the
study of parameter regions where the Mott-Hubbard
 physics ``already'' captured by DMFT becomes important.  Technically,
a full treatment of this regime requires an improvement of the frequency parametrization of the 1PI vertex in the fRG-flow beyond the
current frequency decomposition \cite{Karrasch2008}.

{\sl Summary and outlook.} -- We introduced the DMF$^2$RG approach,
which exploits the synergy of {\sl local} DMFT correlations and  {\sl
  non-local} correlations generated by the fRG flow. Applying
DMF$^2$RG to the $2D$ Hubbard model, we find that, due to the
inclusion of all local correlations by the DMFT starting point, the
divergence of the flow for the interaction vertex is pushed to lower
temperatures, where significant non-local corrections to DMFT are
found. At the same time, in the temperature interval where both fRG
and DMF$^2$RG converge, the self-energy results are qualitatively
similar, supporting the results of previous fRG studies at
weak-to-intermediate $U$. Quantitatively, the most visible effect of
DMF$^2$RG compared to fRG consists in a stronger $\mathbf k$-dependence of the self
energy for the considered parameters and a suppression of the
``pseudocritical'' temperature at which the vertex diverges. We
emphasize, finally, the potential of the presented DMF$^2$RG approach
to access the strong-coupling regime, where the Mott-Hubbard physics
captured by DMFT will play a more important role and qualitative
changes in the self-energy results are to be expected. 
 The flexibility of  the DMF$^2$RG scheme and its ability to avoid
 the sign-problem of a direct QMC treatment of non-local physics beyond DMFT
look promising for future, unbiased studies of correlations in realistic multi-band models.

 We thank A. Eberlein, C. Karrasch, D. Kennes, S. Diehl, T. Enss, O. Gunnarsson, C. Honerkamp, and V. Meden for valuable discussions. 
We acknowledge financial support from FWF through the project I-597-N16,
(DFG research unit FOR 1346, CT) and I-610-N16 (GR, AT), from DFG through research unit FOR 723 (SA, WM) and through grant no. BA 4371/1-1 (JB), from
RFBR grant no. 10-02-91003-ANF\_a  and grant of Dynasty foundation (AK), and FWF SFB ViCoM F41
(KH, SA). Calculations have been performed on the Vienna Scientific
Cluster (VSC).


\begin{thebibliography}{99}
\bibitem{note_cutoff} We note in passing that the fRG data presented in Fig.~\ref{Fig2} have been obtained with the interaction cutoff, but we have verified that these results are robust w.r.t. the choice of the cutoff.
\end{thebibliography}
\end{document}


\begin{center}
\noindent 
{\large {\bf SUPPLEMENTARY MATERIAL} }\\
\vspace{.6em} 

{\large{\bf From infinite to two dimensions through the functional renormalization group}}\\
\vspace{.6em} 

{C.~Taranto, S.~Andergassen, J.~Bauer, K.~Held, A.~Katanin, W.~Metzner, G.~Rohringer, A.~Toschi}

\end{center}
\vspace{4mm}
\noindent 
$\bullet$ {\bf Derivation of the DMF$^2$RG flow equations}

\vspace{2mm}
\noindent
Let us start by  briefly recapitulating the standard fRG technique, which makes
it easier to clarify how the DMFT algorithm can be combined with it. 
 We consider an interacting problem
\begin{eqnarray}
 \mathcal{S}_{\mathrm{latt}} =-\int_0^\beta\!\!d\tau d\tau'
\sum_{\mathbf{k}\sigma}
c^\dag_{\mathbf{k}\sigma}(\tau)G^{0}_{\mathrm{latt}}(\mathbf{k},\tau-\tau')^{-1}c_{\mathbf{k}\sigma}(\tau') + \;\mathcal{S}_{int},
\end{eqnarray}
where $G^0_{\mathrm{latt}}$ is the
propagator of the Gaussian part, 
and all the terms beyond the Gaussian one  are contained in $\mathcal{S}_{\rm int}$; for the further notation  see the paper. 
  In general, the fRG procedure can be  summarized conceptionally in three steps\cite{fRGreview, AiPPHT}:
\begin{enumerate}  
\item  First a ``solvable'' action (${\mathcal S}_{\rm ini}$) is introduced as initial starting point. Here, ``solvable''
  means that at the beginning  the ``problematic'' parts of
  the original action are excluded (e.g., the degrees of freedom close
  to the Fermi level).  Note that ${\mathcal S}_{\rm ini}$  differs
  from the original ${\mathcal S}_{\mathrm{latt}}$ only in its Gaussian part.
 
\item  A one-parameter family of actions  $\mathcal{S}^\Lambda$ is
  defined. These actions smoothly interpolate between the solvable
  action for the initial value of the parameter (i.e., if $\Lambda =
  \Lambda_{\rm ini}$, $\mathcal{S}^{\Lambda_{\mathrm{ini}}}\equiv\mathcal{S}_{\mathrm{ini}}$ )  and the physical one at the end (for $\Lambda =
  \Lambda_{\rm fin}$: $\mathcal{S}^{\Lambda_{\mathrm{fin}}}\equiv\mathcal{S}_{\mathrm{latt}}$).
This corresponds to a continuous change of the Gaussian propagator from $\mathcal{S}_{\rm ini}$ to $\mathcal{S}_{\mathrm{latt}}$. 
 
\item  The  evolution of all (1PI) $m$-particle vertex functions of the actions
  $\mathcal{S}^\Lambda$ as a function of $\Lambda$  is determined from a set of coupled differential equations, called
  ``flow equations''. 
\end{enumerate} 

The formal derivation of this procedure, as well as of
the flow equations for the vertex functions is presented exhaustively  in the literature, see, e.g., the recent reviews Refs.\
\onlinecite{fRGreview,AiPPHT}.

By integrating this set of differential equations, one can {\em in principle} evaluate exactly all 1PI $m$-particle vertex
functions of the action $\mathcal{S}$ of the problem of
interest by computing the flow from the corresponding vertex functions of the solvable model,
independently on which specific choice was made for it.
 However, in the presence of a two-particle interaction, the hierarchy of flow equations couples the $m$-particle
 vertex function $\Gamma^\Lambda_m$ with the $(m+1)$-particle one, i.e., the set of flow
 equations is in general {\sl infinite}. 
Hence, in practice one needs to truncate the equations: As an approximation, it is assumed
that all the 1PI-vertex functions with $m$ bigger than some value
(typically $m_{max}=2$)
are neglected. Within this approximate treatment,  the choice of
the initial action becomes obviously important. 
 
 More specifically, by retaining only the one-particle vertex function
 (self energy) and the two-particle vertex,
 and setting the three-particle vertex to zero, the truncated flow equations assume the form:        
\begin{eqnarray}
\label{eq:flow}
\partial_\Lambda \Sigma^\Lambda&=&\Gamma_2^\Lambda \circ S_\Lambda, \\ 
 \partial_\Lambda \Gamma_2^\Lambda&=& \Gamma_2^\Lambda \circ ( S_\Lambda\circ G_\Lambda) \circ \Gamma_2^\Lambda \;. \label{eq:flow2} 
\end{eqnarray}
Here the symbol ``$\circ$'' stands for the standard summation
 over all internal variables, i.e.,
momentum integration as well as
spin and Matsubara frequency summation. 
 At each vertex, energy, spin, and momentum is conserved according to the conventional diagrammatic rules. 
The symbols $\Sigma^\Lambda$, $\Gamma_2^\Lambda$, $G_\Lambda$ and
$S_\Lambda$ stand respectively for the self energy,
 two-particle vertex, dressed Green's function and single scale
 propagator, as defined in the main text. The initial condition for these differential equations,   $\Sigma^{\Lambda_{\mathrm{ini}}}$,  $\Gamma_2^{\Lambda_{\mathrm{ini}}}$ are obtained 
by solving the initial ``solvable'' action ${\mathcal S}_{\mathrm{ini}}$.
We note, finally, that the Eqs.\ (\ref{eq:flow}) and (\ref{eq:flow2}) correspond diagrammatically to the ones reported in Fig.\ 1 of the manuscript.
Their explicit expression in terms of frequency, momenta and spin
summations can be found, e.g., in Refs. \onlinecite{fRGreview} and 
\onlinecite{Salmhofer2001}.

Let us note that since the fRG flow-equations can not be solved without a truncation (at a given loop level), 
the final results depend on the starting point. In this respect, any fRG flow is ``biased towards'' its starting point. The quality of the results depend on how well the fRG flow can build up the missing part of the physics.  

The basic idea of the new DMF$^2$RG scheme can be summarized as follows:
differently from the conventional fRG approach, in DMF$^2$RG we aim at
including a major part of the correlated physics {\sl already}
 at the level of the initial ``solvable'' action. This is certainly
possible for the non-perturbative, but purely local, correlations of
DMFT,  because the DMFT solution of several models and realistic problems
of solid state physics can be obtained both at the one and the
two-particle level.\cite{Park2011,Rohringer2012,Hafermann2013}  

The formal implementation of this idea requires evidently to replace the
initial action with a one describing the  non-perturbative
local physics of the DMFT solution and then to set up the flow 
 to the final action $\mathcal{S}_{\mathrm{latt}}$ of the desired problem (where {\sl all} correlations, namely also those {\sl beyond} DMFT, are eventually included).
Due to the flexibility of the fRG scheme, there are several ways
to do this in practice. From a mathematical point of view, as
 DMFT corresponds to the exact solution of a quantum many body
Hamiltonian in the limit of infinite dimensions ($d  \rightarrow
\infty$)\cite{DMFT}, the most intuitive way might be realized by building up a 
``dimensional''
flow from $d = \infty$ to the actual dimensions (e.g., $d=2$ or $3$) of the
problem of interest. In this case, one would start from the action of
an infinite dimensional lattice (e.g., hypercubic) and the
  parameter $\Lambda$ should gradually
turn off the  hopping in all directions, except the physical ones 
of the final problem.

This can be done considering the family of actions associated with the following Hamiltonians in the limit $d\rightarrow\infty$ : 
\begin{eqnarray}
\nonumber
 H^\Lambda&=&\sum_{\mathbf{k}\sigma}\left\{\frac{1}{\sqrt{2d}}\left[f(\Lambda)\epsilon_{k_1k_2}+\Lambda\epsilon_{k_3...k_d}\right]-\mu\right\}c_{\mathbf{k}\sigma}^\dag c^{\phantom{\dag}}_{\mathbf{k}\sigma} \\ && +U\sum_i n_{i\uparrow}n_{i\downarrow}. 
\end{eqnarray}

Here the momenta in the first sum are $d$ dimensional: $\mathbf{k}=(k_1,k_2,...,k_d)$ while the second sum extends over the lattice sites of a $d$ dimensional lattice. The operators $c^\dag_{\mathbf{k}\sigma}$ ($c^{\phantom{\dag}}_{\mathbf{k}\sigma}$) create (annihilate) a fermion of momentum $\mathbf{k}$ and spin $\sigma$, $n_{i\sigma}$ is the number operator counting the fermions of spin $\sigma$ at the lattice site $i$. 

To be specific, let us assume that the energies $\epsilon_{k_1k_2}$ refer to a two dimensional square lattice with nearest neighbors hopping $t$: $\epsilon_{k_1k_2}=-2t(\cos k_1+\cos k_2)$. 
The infinite dimensional limit of this lattice is obtained when $\Lambda=1$ and $f(\Lambda)=1$ taking $\epsilon_{k_3...k_d}=-2t(\cos k_3+...+\cos k_d)$. The factor $\frac{1}{\sqrt{2d}}$ accounts for the proper scaling: it guarantees that the kinetic energy does not diverge in the limit $d\rightarrow\infty$. 
The terms $\Lambda$ and $f(\Lambda)$ are used to interpolate between the Hamiltonian in $d$ and two dimensions. For example, assuming the following form for $f(\Lambda)$:

\begin{equation}
 f(\Lambda)=1+(1-\Lambda)(\sqrt{2d}-1),
\end{equation}
one recovers the $d\rightarrow\infty$ limit for $\Lambda=1$, while for $\Lambda=0$ one restores the two dimensional lattice Hamiltonian.

In spite of its intuitive picture, however, such ``dimensional'' flow equations,
 might be not the most suitable scheme to be adopted in practice.
In fact, one should consider that in most of its applications,
and in particular in those aiming at the realistic description of materials, 
DMFT is employed as an ``approximation'' for describing the local physics of
a given {\sl finite}-dimensional system, and no limit of infinite dimensions 
is actually taken. In fact, it would be rather cumbersome
to define a rigorous and general procedure for connecting on a Hamiltonian level, case by case, an infinite dimensional action to a given realistic problem, whose Gaussian part is usually much more complicate than the (hyper)cubic  lattice with nearest-neighbor
hopping.  In fact an effective and flexible implementation of the DMF$^2$RG can be obtained in the formalism of the effective action by including at the beginning of the fRG-flow the local correlated physics obtained from the standard
 application of DMFT as an approximation for the specific finite dimensional (realistic) problem under consideration.  
Formally, this can be achieved in DMF$^2$RG simply by requiring that ${\mathcal S^{\Lambda_{\mathrm{ini}}}}$ is defined by the  action of the auxiliary Anderson impurity model (AIM) associated to the DMFT (self-consistent) solution of the specific, finite dimensional, problem of interest. In practice, this corresponds (i) to calculate the solution of the desired quantum many-body problem  using DMFT as an approximation, then (ii) to extract the 1PI one- [$\Sigma_{\rm DMFT}(i\omega)$] and two-particle [$\Gamma_{\rm DMFT}(i\nu_1,i\nu_2,i\nu'_1,i\nu'_2)$] vertex functions of the associated auxiliary AIM and finally (iii) to use them as initial conditions  of the fRG flow-equations (\ref{eq:flow}), (\ref{eq:flow2}) for the self energy and the vertex functions.  
This way the local correlated physics captured by DMFT will be present from the beginning of the flow, and local and non-local corrections to it will be generated unbiasedly in all channels by the fRG algorithm, via the numerical solution of the associated differential equations. 
For instance at half filling the starting point of the DMF$^2$RG flow, unlike the standard fRG weak coupling starting point, is not only exact in the uncorrelated limit ($U/t\rightarrow 0$), but also in the opposite limit of local atomic physics.
For $U/t \gg 1$, we expect the Mott insulating DMFT solution still to be a good starting point for the fRG, as long as the non-local part of the two particle vertex does not get large.

\vspace{4mm}

\noindent
$\bullet$ {\bf Cutoff scheme of the DMF$^2$RG}

\vspace{2mm}

\noindent
As specified in the previous section, after defining the initial and the final actions, one must also set up a collection of one-parameter dependent actions smoothly interpolating between them. 
In the case of DMF$^2$RG, a quite natural choice is a linear interpolation of the Gaussian part ($G_\Lambda^0(\mathbf{k},i\omega)^{-1}$) of the action from ${\mathcal S^{\Lambda_{\mathrm{ini}}}} = {\mathcal S_{\rm DMFT}}$ (where 
$
 \mathcal{S}_{\mathrm{DMFT}} =-\!\int_0^\beta\! \!d\tau d\tau' \sum_{i\sigma} \bar{c}_{i\sigma}(\tau) \mathcal{G}_{\mathrm{AIM}}^0(\tau-\tau')^{-1}c_{i\sigma}(\tau') 
+ \; \mathcal{S}_{\rm int} 
$)
 to ${\mathcal S^{\Lambda_{\mathrm{fin}}}} = {\mathcal S}_{\mathrm{latt}}$, which reads explicitly:
\begin{equation}
G_\Lambda^0(\mathbf{k},i\omega)^{-1}=f(\Lambda)
  \mathcal{G}_{\mathrm{AIM}}^0 (i\omega)^{-1}+ [1-f(\Lambda)] G^0_{\mathrm{latt}}(\mathbf{k},i\omega)^{-1},
\label{eq:cutoff}
\end{equation}
where $f(\Lambda)$ is an arbitrary smooth function of $\Lambda$ such that $f(\Lambda_{\mathrm{ini}})= 1$ and $f(\Lambda_{\mathrm{fin}})= 0$. For the sake of clarity, in the manuscript we have chosen $f(\Lambda)= \Lambda$ with $\Lambda_{\mathrm{ini}}=1$ and  $\Lambda_{\mathrm{fin}}=0$, but, obviously, any alternative choice of $f(\Lambda)$ simply leads to an equivalent formulation of the truncated flow equations. 
 
This choice of ``cutoff'' scheme  ${\mathcal S^{\Lambda}}$ according to Eq. (\ref{eq:cutoff}) is similar to the ``interaction cutoff''\cite{HRAE2004} in the standard fRG, since it does not operate any selective cut on specific regions of the momentum and/or frequency space. 
The implementation of a frequency cutoff, may read
\begin{equation}
G_\Lambda^0(\mathbf{k},i\omega)^{-1}= \theta(\Lambda-|\omega|) {\mathcal G}_{\mathrm{AIM}}^0(i\omega)^{-1} + \theta(|\omega|-\Lambda) G^0_{\mathrm{latt}}(i\omega,{\mathbf k})^{-1},
\label{eq:freqcutoff}
\end{equation}
where $\theta(x)$ is the Heavyside-step function. Evidently all possible cutoff schemes are equivalent in the case of a non-truncated flow. In the actual implementation however, a frequency- or momentum-cutoff, which can regularize infrared divergences of the problem, might be more suited, in particular, to study the regime in the proximity of (quantum) phase transitions. Its effective implementation, however, is numerically more involved than the simple cutoff of Eq.\ (\ref{eq:cutoff}) and  subject to future investigations.

\vspace{4mm}
\noindent 
$\bullet$ {\bf Parametrization of the two-particle vertex}
\vspace{2mm}

\noindent

\begin{figure}[tb] 
\begin{center} 
\includegraphics[width=15cm, clip=true]{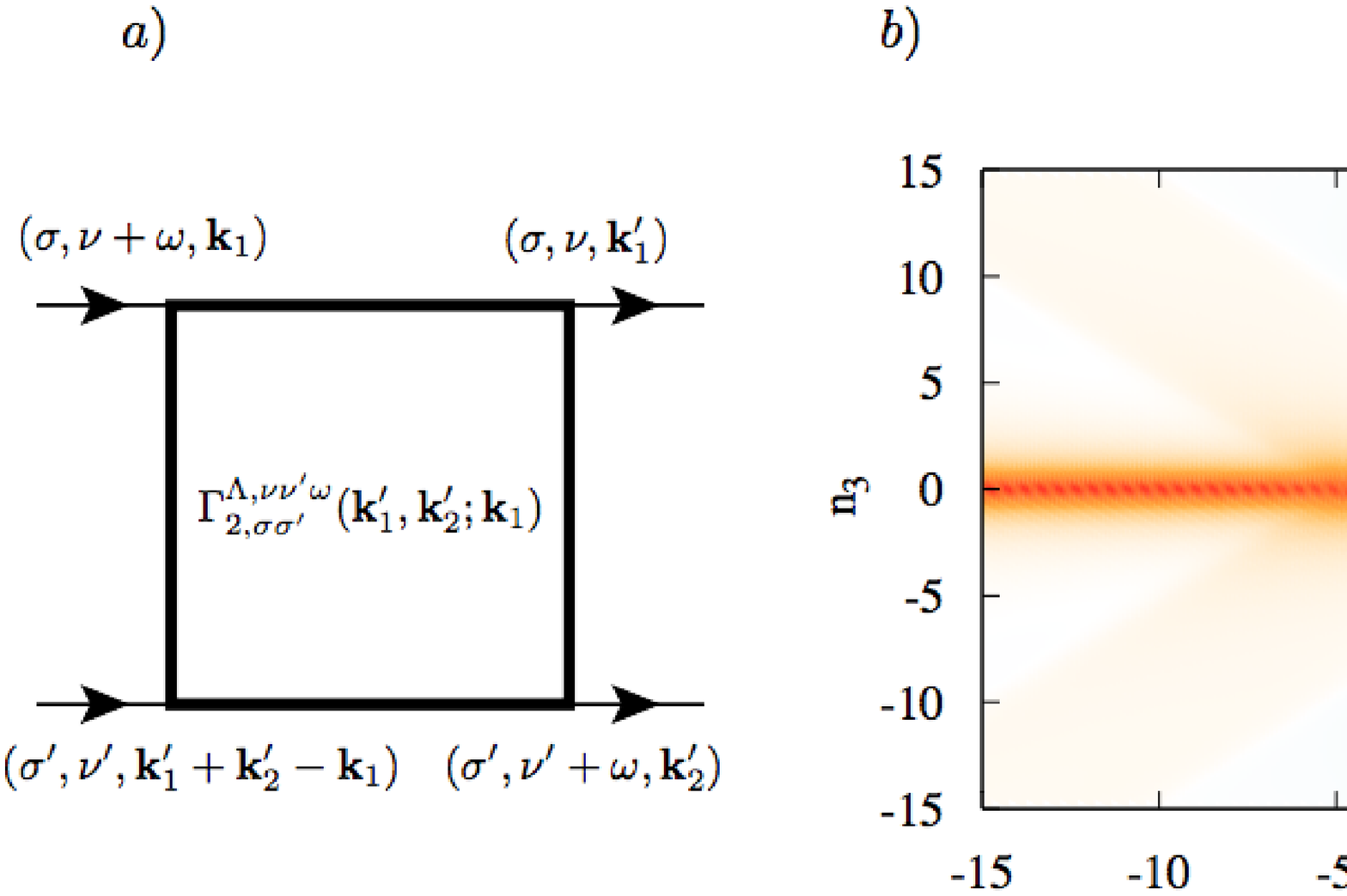} 
\vspace{-4.0cm} 

\end{center} 
\caption{(Color online) On the left, diagrammatic representation of the two-particle 1PI vertex function $\Gamma_{2}^\Lambda$; the arrows mark the position of the incoming and outgoing electrons. On the right, DMFT vertex function $\Gamma_{2,\uparrow\downarrow}^{\Omega_1\Omega_2\Omega_3}$  as a function of $\Omega_1$ and $\Omega_3$ at fixed $\Omega_2$ ($\Omega_i=2\pi T n_i$; $n_2=20$) for the Hubbard model on a three dimensional cubic lattice with nearest neighbors hopping and $U=0.5D$, $T=0.038D$ ($D$ being the half bandwidth). The color coded values are measured in units of $D$. 
Please note that in DMFT the main features of the vertex {\em do not} depend on the details of the lattice, but only on the bandwidth, therefore the following considerations apply in general. 
The white background color corresponds to the asymptotic value $U$ reached by the vertex. 
On the top of this, one can recognize three structures: $i$) a vertical line at $\Omega_1=0$, $ii$) a horizontal line at $\Omega_3=0$, and $iii$) a broader (hardly discernible) cross structure on the diagonals at $\Omega_1=\pm\Omega_3$.
The origin of the three structures has been analyzed in Ref. \onlinecite{Rohringer2012}. While the structures $i$) and $ii$) are well described by the frequency dependence approximation described in the text, the cross structure is not captured by the approximation. Please notice that the white corners on the right of the density plot correspond to frequencies not included in the frequency window of our data set.}  
\label{vertex}
\end{figure}
Here we give some details about the approximation employed on the frequency dependence of the 1PI two-particle vertex function for the single-band Hubbard model, studied in the manuscript.
We use the conventions and definitions of Ref. \onlinecite{Rohringer2012} (in particular in ``{\sl particle-hole notation}''). Please note, however, that there the two-particle 1PI vertex is labeled $F$ and is momentum independent, while here it is called $\Gamma_2^\Lambda$ and can depend on the momentum.

 In general for an SU(2) symmetric interaction and for a translationally invariant system the vertex function depends on two spins, three frequencies, and three momenta variables (see Fig. \ref{vertex}a):
\begin{equation}
 \Gamma_{2,\sigma\sigma'}^{\Lambda,\nu\nu'\omega}(\mathbf{k}'_1,\mathbf{k}'_2;\mathbf{k}_1):=\Gamma_{2}^\Lambda(\phantom{1}\underbrace{\nu\mathbf{k}'_1\sigma,(\nu'+\omega)\mathbf{k}'_2\sigma'}_{\mathrm{outgoing\phantom{1} electrons}}\phantom{1};\phantom{1}\underbrace{(\nu+\omega)\mathbf{k}_1\sigma,\nu'(\mathbf{k}'_1+\mathbf{k}'_2-\mathbf{k}_1)\sigma'}_{\mathrm{incoming\phantom{1} electrons}}\phantom{1}).
\end{equation}
Here $\nu$ and $\nu'$ are fermionic Matsubara frequencies, while $\omega$ is a bosonic Matsubara frequency. 
Physically this describes the scattering of a hole of energy $-\nu$ with an electron of energy $\nu+\omega$.
 
For the implementation of the frequency parametrization used in previous fRG studies\cite{Karrasch} it is  however more convenient to adopt a frequency notation in terms of three \emph{bosonic} Matsubara frequencies\cite{Karrasch} defined as follows:
\begin{eqnarray}
\Omega_1&=& \nu+\nu'+\omega, \\
\Omega_2 &=& \nu-(\nu+\omega)=-\omega, \\
\Omega_3 &=& \nu'+\omega-(\nu+\omega)=\nu'-\nu.
\end{eqnarray}

As for the spin indexes by exploiting the SU(2) symmetry we have
\begin{equation}
 \Gamma_{2,\uparrow\uparrow}^{\Lambda,\Omega_1\Omega_2\Omega_3}(\mathbf{k}'_1,\mathbf{k}'_2;\mathbf{k}_1)=\Gamma_{2,\uparrow\downarrow}^{\Lambda,\Omega_1\Omega_2\Omega_3}(\mathbf{k}'_1,\mathbf{k}'_2;\mathbf{k}_1)-\Gamma_{2,\uparrow\downarrow}^{\Lambda,\Omega_1\Omega_3\Omega_2}(\mathbf{k}'_2,\mathbf{k}'_1;\mathbf{k}_1).
\end{equation}
Hence, we can concentrate on the vertex $\Gamma^\Lambda_{2,\uparrow\downarrow}$ only (all the other spin combinations can be obtained by symmetry).

Even by restricting ourselves to the $\uparrow\downarrow$ sector, the vertex function $\Gamma^\Lambda_{2,\uparrow\downarrow}$ displays, in general, a rather complicated structure in momentum and frequency space. 
However, the efficiency of our first DMF$^2$RG calculations could be improved by parametrizing the frequency dependence of $\Gamma^2_\Lambda$ following the previous experience of an fRG study of the AIM\cite{Karrasch}, where the following approximation for the frequency dependence of the vertex: 
\begin{equation}
\label{decomposition}
 \Gamma_{2,\uparrow\downarrow}^{\Lambda\Omega_1\Omega_2\Omega_3}(\mathbf{k}'_1,\mathbf{k}'_2;\mathbf{k}_1)\approx U+\tilde\Gamma_{2,\mathrm{pp}}^{\Lambda,\Omega_1}(\mathbf{k}'_1,\mathbf{k}'_2;\mathbf{k}_1)+\tilde\Gamma_{2,\mathrm{ph-d}}^{\Lambda,\Omega_2}(\mathbf{k}'_1,\mathbf{k}'_2;\mathbf{k}_1)+\tilde\Gamma_{2,\mathrm{ph-c}}^{\Lambda,\Omega_1}(\mathbf{k}'_1,\mathbf{k}'_2;\mathbf{k}_1),
\end{equation}
is proposed.

This corresponds to approximating the complicated dependence of $\Gamma_2^\Lambda$ on the three bosonic frequencies $\Omega_1$, $\Omega_2$, $\Omega_3$, assuming that the scattering amplitude among two particles can be completely decomposed in three different channels ($\mathrm{pp}$, $\mathrm{pp-d}$, $\mathrm{ph-c}$).
This assumption is not exact for a generic $U$, as one could immediately see already by looking at the DMFT vertex function, which represents the input for DMF$^2$RG. However, as described in detail in Refs. \onlinecite{Karrasch,Rohringer2012}, is consistent with the lowest-order perturbation theory for $\Gamma_{2,\uparrow\downarrow}^\Lambda$. 
Following Ref. \onlinecite{Karrasch} one can derive the flow equations directly for the functions $\tilde\Gamma_{2,\mathrm{x=pp,ph-d,ph-c}}^\Lambda$. This is possible because one can associate each function with a specific channel: particle-particle, particle-hole direct and particle-hole crossed.

In fact, it has been shown that this approximation correctly describes the main vertex structures up to $\mathcal{O}(U^3)$ and it is expected to be reliable for moderate values of $U$.
On the other hand, increasing the $U$ value, structures not captured by Eq. (\ref{decomposition}), and arising from higher order diagrams (like the diagonal ones at $\Omega_1=\pm\Omega_3$ in Fig. \ref{vertex}b) will become more important, making the approximation unreliable. This is the reason why we only present results for moderate $U$ values in the paper.

The only point left to discuss is how to extract the initial condition for the three functions in Eq. (\ref{decomposition}) from the fully frequency dependent DMFT vertex $\Gamma_{\mathrm{DMFT}}^{\Omega_1\Omega_2\Omega_3}$ which contains more information than necessary. By looking at Fig. \ref{vertex}b, one sees that the problem consists in how to get rid of the cross structure (labeled $iii$) in the caption of Fig. \ref{vertex} which depends on all the frequencies. However, the structure under consideration fades out becoming gradually broader and less intense as the third frequency is increased.
Therefore to extract one of the three functions, say, e.g., $\tilde\Gamma_{2,\mathrm{pp}}^{\Lambda_{\mathrm{ini}},\Omega_1}$, it suffices to take a cut in $\Gamma_{\mathrm{DMFT}}$ keeping $\Omega_2$ and $\Omega_3$ fixed at some very large values $\Omega_2^c$ and $\Omega_3^c$: 
\begin{equation}
 \tilde\Gamma_{2,\mathrm{pp}}^{\Lambda_{\mathrm{ini}},\Omega_1}(\mathbf{k}'_1,\mathbf{k}'_2;\mathbf{k}_1)=\Gamma_{\mathrm{DMFT}}^{\Omega_1\Omega_2^c\Omega_3^c}.
\end{equation}